\documentclass[a4paper,10pt]{article}

\pdfoutput=1
\usepackage{fullpage}
\usepackage[utf8]{inputenc}
\usepackage{graphicx}
\usepackage{amssymb}
\usepackage{amsmath}
\usepackage{amsfonts}
\usepackage{color}
\usepackage[pdftex]{hyperref}
\usepackage[english]{babel}
\usepackage{algorithm}
\usepackage{algorithmic}
\usepackage{tabularx,colortbl}
\usepackage[sort&compress,numbers]{natbib}
\usepackage{authblk}
\usepackage[title]{appendix}
\usepackage{array}
\usepackage{multirow}
\usepackage{lineno}
\usepackage{subfigure}
\usepackage{sidecap}
\sidecaptionvpos{figure}{c}

\definecolor{Orange}{rgb}{1,0.64,0}
\newcommand{\argmin}{\operatornamewithlimits{arg\ min}}
\newcommand{\vect}[1]{\mathbf{#1}}
\DeclareMathSymbol{\R}{\mathalpha}{AMSb}{"52}
\newcommand{\norm}[2][2]{\left\lVert #2 \right\rVert_{#1}} 

\makeindex

\begin{document}

\title{Data-driven detrending of nonstationary fractal time series with echo state networks}

\author[1]{Enrico Maiorino\thanks{enrico.maiorino@uniroma1.it}}
\author[2]{Filippo Maria Bianchi\thanks{filippomaria.bianchi@uniroma1.it}}
\author[3]{Lorenzo Livi\thanks{L.Livi@exeter.ac.uk}\thanks{Corresponding author}}
\author[1]{Antonello Rizzi\thanks{antonello.rizzi@uniroma1.it}}
\author[4]{Alireza Sadeghian\thanks{asadeghi@ryerson.ca}}
\affil[1]{Dept. of Information Engineering, Electronics, and Telecommunications, SAPIENZA University of Rome, Italy}
\affil[2]{Machine Learning Group, Department of Physics and Technology, University of Troms\o{} - The Arctic University of Norway}
\affil[3]{Dept. of Computer Science, College of Engineering, Mathematics and Physical Sciences, University of Exeter, UK}
\affil[4]{Dept. of Computer Science, Ryerson University, Toronto, Canada}
\renewcommand\Authands{, and }
\providecommand{\keywords}[1]{\textbf{\textit{Keywords---}} #1}

\maketitle

\begin{abstract}
In this paper, we propose a novel data-driven approach for removing trends (detrending) from nonstationary, fractal and multifractal time series. We consider real-valued time series relative to measurements of an underlying dynamical system that evolves through time. 
We assume that such a dynamical process is predictable to a certain degree by means of a class of recurrent networks called Echo State Network (ESN), which are capable to model a generic dynamical process.
In order to isolate the superimposed (multi)fractal component of interest, we define a data-driven filter by leveraging on the ESN prediction capability to identify the trend component of a given input time series.
Specifically, the (estimated) trend is removed from the original time series and the residual signal is analyzed with the multifractal detrended fluctuation analysis procedure to verify the correctness of the detrending procedure.
In order to demonstrate the effectiveness of the proposed technique, we consider several synthetic time series consisting of different types of trends and fractal noise components with known characteristics.
We also process a real-world dataset, the sunspot time series, which is well-known for its multifractal features and has recently gained attention in the complex systems field.
Results demonstrate the validity and generality of the proposed detrending method based on ESNs.\\
\end{abstract}

\section{Introduction}
\label{sec:intro}

Memory is one of the most interesting aspects of many processes in Nature and society \cite{gao2007multiscale}. In order to characterize and predict a system with memory, it is necessary to keep into account its past history. Memory can be quantified in different ways, depending on the particular features and effects of interest. One of the most common approaches in the study of real-valued time series is the analysis of the autocorrelation function. In such a linear setting, the extent of memory can be roughly quantified through the decay of the autocorrelation function, which indicates the characteristic time scales at which the series remains correlated.
When the decay is exponential, the series is said to manifest \textit{short-term memory} and the influence of the past to the current state is limited in time.
Instead, if the decay follows a power-law, then there is no characteristic scale in the autocorrelation, i.e., the influences of the past have no cut-off.
In this case, a time series is said to manifest \textit{long-term memory} or \textit{long-term correlation} (LTC) and the strength of this correlation is referred to as degree of \textit{persistence} of the generating stochastic process.
Persistence of a stochastic process \cite{sanchez2014effect} is quantified by the self-similarity coefficient of the process' fluctuations, called Hurst exponent $H\in[0, 1]$. A straightforward numeric approach to estimate the Hurst coefficient is the Fluctuation Analysis (FA), which evaluates the slope of the fluctuations scaling function $F(s)$. This function is in turn calculated by dividing the integrated time series in segments of equal sizes $s$ and evaluating the root mean square difference between their extremal points. When the process corresponds to uncorrelated noise (e.g., white Gaussian noise), then the value of $H$ is 0.5, whereas if the process is persistent (correlated) or antipersistent (anticorrelated) it will be respectively greater than or less than 0.5.

However, conventional methods employed to analyze the LTC properties of a time series (e.g., FA, spectral analysis, R/S analysis \cite{serinaldi2010use,barunik2010hurst,shao2012comparing}) are misleading when such time series are non-stationary \cite{PhysRevE.65.041107}.
In fact, in many cases a process is driven by underlying trends \cite{PhysRevE.64.011114}, which operate at specific time scales, like seasons in the analysis of data related to a natural phenomenon and days in financial market analysis. 
Usually, when investigating memory properties of a process, one is interested in the fractal properties of the intrinsic fluctuations of such a process.
Hence, to analyze the fluctuations of the stationary component of a time series, it is necessary to remove the non-stationary trend components. This can be done by employing one of the several methods proposed for this purpose, like detrended fluctuation analysis (DFA), detrended moving average, wavelet leaders \cite{wendt2007multifractality}, adaptive fractal analysis \cite{riley2012tutorial}, and the so-called geometric-based approaches \cite{fernandez2013measuring}.
Notably, DFA has been shown to be successful in a broad range of applications \cite{kantelhardt2009fractal,shao2012comparing,parkisontremors_mfs2015}.
The DFA has been generalized in the so-called Multifractal Detrended Fluctuation Analysis (MFDFA) \cite{kantelhardt2002multifractal,PhysRevE.74.016103,bashan2008comparison,PhysRevLett.62.1327}, which accounts for the existence of multiple scaling exponents in the same data.
DFA and the related variants remove trends from data by means of window-based local (polynomial) fittings. However, trends are often defined in terms of periodicities and/or fast-varying functions, resulting thus in a spurious detection of fractality \cite{grech2013multifractal}.
For this reason, additional detrending methods are often used as a preprocessing step of the (MF-)DFA to single out these trends before the polynomial detrending takes place. In other research works, the local detrending step of DFA is modified or replaced with other ad-hoc methods \cite{hu2009multifractal,qian2011modified,eaf2015_ijbc}.

The main problem with detrending lies in the difficulty of defining what exactly a trend is \cite{wu2007trend}. Local-fit based methods rely on the assumption that a trend is generally a slow-varying process, while the superimposed noise is a process characterized by higher frequencies.
While this is often the case, it is still difficult to determine the right form and parameters of the fitting function without biasing the analysis.
Moreover, window-based fitting algorithms are heavily influenced by the choice of the window sizes. In \cite{wu2007trend} a trend is defined as an intrinsically fitted monotonic function or a function in which there can be at most one extremum within a given data span. 
This method is not affected by border effects since it is not window-based. However, a problem with this definition is that it does not (fully) describe periodic trends in a consistent way.
\citet{chianca2005fourier} suggested to perform a detrending by applying a simple low-pass filter, in order to eliminate slow periodic trends from data. While this approach is suitable for systems with slow-varying trends, it is difficult to apply to more general cases, when the trends' frequencies span over a significant portion of the (power) spectrum.
Another approach that has been demonstrated to be useful in the case of periodicities was proposed by \citet{nagarajan2006reliable}.
As a first step, the signal is represented as a matrix, whose dimension has to be much larger than the number of frequency components of the periodic (or quasi-periodic) trends as shown by the power spectrum. The well-known singular value decomposition method is then applied to remove components related to large-magnitude eigenvalues, which correspond to the trend. Such a method, although interesting and mathematically well-founded, is very demanding in terms of computations and also assumes a deterministic form for trends.

In this work, we follow an approach similar to \citet{wu2007trend} and define a trend in a completely data-driven way. We consider the analyzed time series as a series of noisy measurements of an unknown dynamical process.
We also assume that the dynamical process is predictable to a certain degree by means of a particular type of Recurrent Neural Network (RNN) called Echo State Network (ESN) \cite{lukovsevivcius2009reservoir, bianchi2015prediction}. RNNs have been shown to be able to predict the outcome of a number of dynamical processes \cite{dambre2012information}. In particular, a fundamental theorem formulated within the Neural Filtering framework, relates the number of neurons in a RNN hidden layer with the expected approximation accuracy of the estimated signal with respect to the true signal \cite{lo1994synthetic} of the process. Specifically, given a sufficiently large amount of processing units, a RNN that takes as input the measurement process can output an estimation that can be made as close as desired to the signal process, given its past input sequences.
However, not all processes are predictable at the same level, as formally studied in \cite{bialek2001predictability,crutchfield2003regularities}, for instance.
For example, chaotic processes are not predictable for long time-steps, while other deterministic systems, like a sinusoidal waveform, can be easily predicted.
In a stochastic setting, instead, we note that white noise cannot be predicted at all, since the past observations do not convey any information about the future.
On the other hand correlated noise signals, such as fractional Gaussian noise (fGn), are in theory partially predictable given the presence of memory in the process.
To handle prediction problems of increasing difficulty, models characterized by a higher complexity or a larger amount of training data are required.
In the case of ESNs, the complexity of the model is mainly determined by the properties and the size of its recurrent hidden layer.
Here we propose to perform a data-driven detrending of nonstationary, fractal and multifractal time series by using ESNs acting as a filter. 
In this study, trends are the only form of nonstationarities that we consider.
By means of ESNs, we predict the trend of a given input time series, which is always superimposed to the (multi)fractal component of interest. Such a trend is then removed from the original time series and the residual signal is analyzed with MFDFA in order to evaluate its scaling and (multi)fractal properties.

The remainder of the paper is structured as follows.
In Section \ref{sec:background}, we provide technical background related the main tools utilized in this work, namely the MFDFA procedure and ESNs.
In Section \ref{sec:ens_trend} we present the detrending method based on ESNs. Notably, we employ an ensemble of ESNs with fixed complexity (i.e., network sizes and connectivities) to separate the (multi)fractal signals from the trends.
In Section \ref{sec:exp}, we show and discuss the experimental results obtained on a number of time series related to mathematical models and on the well-known sunspot data.
Finally, concluding remarks are provided in Section \ref{sec:conclusions}.

\section{Technical background}

In this Section we provide a technical background on the main tools used in this work: Multifractal Detrended Fluctuation Analysis and Echo State Networks.

\label{sec:background}


\subsection{Multifractal detrended fluctuation analysis}
\label{sec:mfdfa}

The Multifractal Detrended Fluctuation Analysis, a generalization of the Detrended Fluctuation Analysis, is at present one of the most efficient methods for estimating and quantifying the fractal properties of a time series. The procedure is described thoroughly in \cite{kantelhardt2002multifractal} and is reported briefly in the following.
Let $\mathbf{x} = \{ x(t) \}_{t=1}^{T}$ be a time series of length $T$ with compact support. The procedure consists of five steps, three of which are identical to the DFA version.

\begin{itemize}
\item{\textit{Step 1} : Evaluate $Y(i)$ as the cumulative sum (profile) of the series as
\begin{equation}
Y(i) \equiv \sum_{t=1}^{i} \left[ x(t) - \langle \mathbf{x} \rangle \right], \;\; i = 1, \dotsc, T.
\end{equation}
}
\item{\textit{Step 2} :
Separate $Y(i)$ in $N_s \equiv \lfloor(T/s)\rfloor$ non-overlapping segments of equal
length $s$, $\lfloor \cdot \rfloor$ being the floor operation. To account for non-zero remainders of the division, this operation is repeated in reverse order starting from the opposite end of the series, thus obtaining a total of $2N_s$ segments.}
\item{\textit{Step 3} : Perform local detrending by fitting a polynomial functional form on
each of the $2N_s$ segments. Then determine the variance,
\begin{equation}
F^2(\nu,s) \equiv \frac{1}{s} \sum_{i=1}^s \bigg\{ Y[(\nu-1)s+i] - y_\nu(i) \bigg\}^2,
\end{equation}
for each segment $\nu = 1,\dotsc,N_s$ and
\begin{equation}
F^2(\nu,s) \equiv \frac{1}{s}\sum_{i=1}^s \bigg\{ Y[T-(\nu - N_s)s+i] - y_\nu(i)\bigg\}^2
\end{equation}
for $\nu = N_s +1,\dotsc, 2N_s$, where $y_\nu(i)$ is the fitted polynomial in segment $\nu$. The order $m$ of the fitting polynomial, $y_\nu(i)$, determines the extent of the (MF-)DFA in filtering out trends in the series, thus it has to be tuned according to the expected maximum trending order of the time series.
}
\item{\textit{Step 4} : Compute the $q$th-order average of the variance over all segments,
\begin{equation}
\label{eq:Fq}
F_q(s) \equiv \bigg\{ \frac{1}{2N_s} \sum_{\nu=1}^{2N_s} \left[ F^2(\nu,s)\right]^{q/2} \bigg\}^{1/q},
\end{equation}
with $q \in \mathbb{R}$. The $q$-dependence of the fluctuations function $F_q(s)$ highlights the contribution of fluctuations at different orders of magnitude.
For $q > 0$ ($q < 0$) only larger (smaller) fluctuations contribute mostly to the average in Eq.~\eqref{eq:Fq}. For $q = 2$ the standard DFA procedure is obtained. The case $q = 0$ is not well defined with the averaging form in Eq.~\eqref{eq:Fq} and so a logarithmic form has to be employed,
\begin{equation}
F_0(s) = \exp \bigg\{ \frac{1}{2N_s} \sum_{\nu=1}^{2N_s} \ln \left[F^2(\nu,s)\right] \bigg\}.
\end{equation}
Steps 2 to 4 are repeated for different time scales $s$, where all values of $s$ have to be chosen such that $s \geq m+2$ to allow for a meaningful fitting of data. It is also convenient to avoid scales $s > T/4$ because of the statistical unreliability of such small numbers $N_s$ of segments considered.
} 
\item{\textit{Step 5} : Determine the scaling behavior of the fluctuation functions by analyzing log-log plots of $F_q(s)$ versus $s$ for each value of $q$. If the series $x_i$ is long-range power-law correlated, $F_q(s)$ is approximated (for large values of $s$) by the form
\begin{equation}
\label{eq:Fqshq}
F_q(s) \sim s^{h(q)}.
\end{equation}} 
\end{itemize}

The exponent $h(q)$ is the \textit{generalized Hurst exponent}; for $q=2$ and stationary time series, $h(q)$ reduces to the standard Hurst exponent, $H$. When the time series manifests a uniform scaling over all magnitudes of fluctuations - i.e. $h(q)$ is independent of $q$ - the series is said monofractal. On the contrary, when $h(q)$ is spread over several values the series is multifractal.

Starting from Eq.~\eqref{eq:Fq} and using Eq.~\eqref{eq:Fqshq}, it is straightforward to obtain
\begin{equation}
\sum_{\nu=1}^{T/s} [ F(\nu,s)]^q \sim s^{qh(q) - 1},
\end{equation}
where, for simplicity, it has been assumed that the length $T$ of the series is a multiple of the scale $s$, such that $N_s = T/s$.
The exponent
\begin{equation}
\label{eq:tauq}
\tau(q) = qh(q) - 1
\end{equation}
corresponds to the multifractal generalization of the \textit{mass exponent}. In case of positive stationary and normalized time series, $\tau(q)$ corresponds to the scaling exponent of the $q$-order partition function $Z_q(s)$.
Another quantity that characterizes the multifractality of a series is the \textit{singularity spectrum}, $D(\alpha)$, which is obtained by applying the Legendre transform to $\tau(q)$,
\begin{equation}
\label{eq:mutifractal_spectrum}
D(\alpha) = q\alpha - \tau(q),
\end{equation}
where $\alpha$ is equal to the derivative $\tau'(q)$ and corresponds to the \textit{H\"older} or \textit{singularity exponent}. Using Eq.~\eqref{eq:tauq} it is possible to directly relate $\alpha$ and $D(\alpha)$ to $h(q)$, obtaining:
\begin{equation}
\alpha = h(q) + qh'(q) \;\; \text{and} \;\; D(\alpha) = q[\alpha - h(q)] + 1.
\end{equation}

The multifractal spectrum in Eq. \eqref{eq:mutifractal_spectrum} allows to infer important information regarding the ``degree of multifractality'' and the specific sensitivity of the time series to fluctuations of different magnitudes.
In fact, the width of the support of $D(\cdot)$ is an important quantitative indicator of the multifractal character of the series (the larger, the more multifractal a series is).
Also the codomain of $D(\cdot)$ encodes useful information, since it corresponds to the dimension of the subset of the times series domain which is characterized by the singularity exponent $\alpha$.


\subsection{Echo state networks}
\label{sec:esn}

\begin{SCfigure}[1.2][!ht]
	\centering
	\includegraphics[width=0.35\textwidth, keepaspectratio]{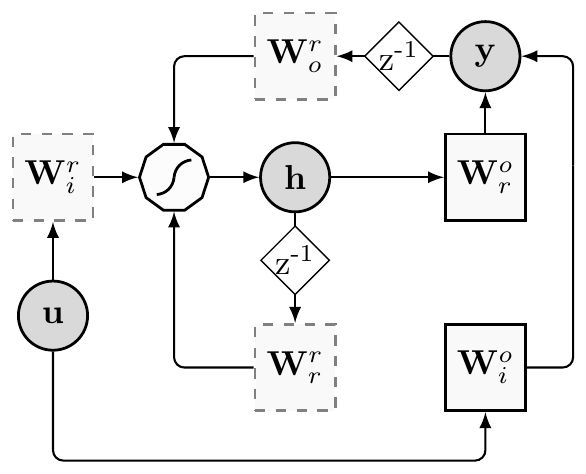}
	\caption{Schematic depiction of an ESN. The circles represent the input variables $\mathbf{u}$, the state variables $\mathbf{h}$ and the output variables $\mathbf{y}$. The squares depicted with solid lines, $\mathbf{W}_{r}^{o}$ and $\mathbf{W}_{i}^{o}$, are the trainable weight matrices of the readout, while the squares with dashed lines, $\mathbf{W}_{r}^{r}$, $\mathbf{W}_{o}^{r}$ and $\mathbf{W}_{i}^{r}$, are random initialized weight matrices. The polygon represents the non-linear transformation performed by neurons and $\text{z}^{\text{-1}}$ is the backshift/lag operator.}
	\label{fig:esn}
\end{SCfigure}

ESNs belong to the class of computational dynamical systems, implemented according to the biologically-inspired reservoir computing approach \cite{lukovsevivcius2009reservoir}. An input signal is fed to a large, recurrent and randomly connected hidden layer, the reservoir, whose outputs are combined by a memory-less linear layer, called readout, to solve a specified task.
ESNs have been adopted in a variety of different contexts, such as time series prediction \cite{7286732}, static classification \cite{alexandre2009benchmarking}, speech
recognition \cite{skowronski2007automatic}, adaptive control \cite{6480841} harmonic distortion measurements \cite{4712533} and, in general, for modeling of various kinds of non-linear dynamical systems \cite{han2014fuzzy}.
A schematic depiction of an ESN is shown in Fig. \ref{fig:esn}. 

The current output of an ESN is computed in two distinct phases. First, the $N_i$-dimensional input vector $\vect{u}(t) \in \R^{N_i}$ is given as input to the recurrent reservoir, whose internal state $\vect{h}(t-1) \in \R^{N_r}$ is updated according to the state equation:
\begin{equation}
	\vect{h}(t) = f_{\text{res}}\left(\vect{W}_i^r \vect{u}(t) + \vect{W}_r^r \vect{h}(t-1) + \vect{W}_o^r \mathbf{y}(t-1)\right),
	\label{eq:esn_state_update}
\end{equation}
where $\vect{W}_i^r \in \R^{N_r \times N_i}$, $\vect{W}_r^r \in \R^{N_r \times N_r}$ and $\vect{W}_o^r \in \R^{N_r \times N_o}$ are randomly initialized at the beginning of the learning process, and they remain unaltered afterwards. $f_{\text{res}}(\cdot)$ in Eq. \eqref{eq:esn_state_update} is a suitable non-linear function, typically a sigmoid, and $\vect{y}(t-1) \in \R^{N_o}$ is the previous output of the network. In our case, we have $f_{\text{res}}(\cdot) = \tanh(\cdot)$. In the second phase, the ESN prediction is computed according to:
\begin{equation}
\label{eq:esn_output_update}
\mathbf{y}(t) = \vect{W}_i^o \vect{u}(t)  + \vect{W}_r^o \vect{h}(t) \,,
\end{equation}
where $\vect{W}_i^o \in \R^{N_o \times N_i}, \vect{W}_r^o \in \R^{N_o \times N_r}$ are trainable connections. The difference between fixed and adaptable weight matrices is shown in Fig. \ref{fig:esn} with the use of continuous and dashed lines, respectively. 

Finally, a few words should be spent on the choice of the matrix $\vect{W}_r^r$. According to the ESN theory, the reservoir must satisfies the so-called ``echo state property'' (ESP) \cite{lukovsevivcius2009reservoir}. This means that the effect of a given input on the state of the reservoir must vanish in a finite number of time-instants. In this paper we adopt the widely used rule-of-thumb that suggests to rescale the matrix $\vect{W}_r^r$ to have $\rho(\vect{W}_r^r) < 1$, where $\rho(\cdot)$ denotes the spectral radius.

To determine the weight matrices of the readout, let us consider a training sequence of $T_\text{tr}$ desired input-outputs pairs 
$\left\{ \vect{u}(t), \mathbf{d}(t) \right\}_{t=1}^{T_\text{tr}} $, where the output is given by $\mathbf{d}(t) = \mathbf{u}(t+\tau_f)$.
Here, $\tau_f$ defines the forecast horizon (or step ahead) considered in the prediction, i.e. how far ahead in time the input signal must be predicted. In the initial phase of training, called ``state harvesting'', the inputs are fed to the reservoir in accordance with Eq. \eqref{eq:esn_state_update}, producing a sequence of internal states $\{ \vect{h}(t) \}_{t=1}^{T_\text{tr}}$. Since, by definition, the outputs of the ESN are not available for feedback, the desired output is used instead in Eq. \eqref{eq:esn_output_update} (the so-called ``teacher forcing''). The states are stacked in a matrix $\vect{H} \in \R^{T_\text{tr} \times N_i + N_r}$ and the desired outputs in a vector $\vect{d} \in \R^Q$:
\begin{align}
\vect{H} = & 
\left[\begin{array}{c}
\vect{u}^T(1), \,\, \vect{h}^T(1) \\
\vdots \\
\vect{u}^T(T_\text{tr}), \,\, \vect{h}^T(T_\text{tr})
\end{array}\right] \label{eq:state_matrix} \,,\\
\vect{d} =  &
\left[\begin{array}{c}
d(1) \\
\vdots \\
d(T_\text{tr})
\end{array}\right] \,.\label{eq:output_vector}
\end{align}
The initial $D$ rows from Eq. \eqref{eq:state_matrix} and Eq. \eqref{eq:output_vector} should be discarded, since they refer to a transient phase in the ESN's behavior. We refer to them as the washout elements.

At this point the resulting training problem is a standard linear regression, which can be solved in a large variety of ways. We used the least-square regression, which is the algorithm originally proposed for training the readout \cite{jaeger2010the}. It consists in the following regularized least-square problem:
\begin{equation}
\vect{w}_{\text{ls}}^* = \argmin_{\vect{w} \in \R^{N_i+N_r}}  \frac{1}{2}\norm{\vect{H}\vect{w} - \vect{d}}^2 + \frac{\alpha}{2}\norm{\vect{w}}^2 \,,
\label{eq:esn_opt}
\end{equation}
where $\vect{w} = \left[ \vect{w}_i^o \, \vect{w}_r^o \right]^T$ and $\alpha \in \R^+$ is a positive scalar known as \textit{regularization factor}. A solution of problem \eqref{eq:esn_opt} can be obtained in closed form as:
\begin{equation}
\vect{w}_{\text{ls}}^* = \left( \vect{H}^T\vect{H} + \alpha \vect{I} \right)^{-1}\vect{H}^T \vect{d} \,.
\label{eq:ridge_regression_minimizer}
\end{equation} 
Whenever $N_r + N_i > Q$, Eq. \eqref{eq:ridge_regression_minimizer} can be computed more efficiently by rewriting it as:
\begin{equation}
\vect{w}_{\text{ls}}^* = \vect{H}^T\left( \vect{H}\vect{H}^T + \alpha \vect{I} \right)^{-1}\vect{d} \,.
\label{eq:ridge_regression_minimizer_2}
\end{equation} 

Once the readout layer is trained, when the network is fed with an unseen input signal $\mathbf{u}(t)$, with $t > T_\text{tr}$, it returns a predicted value $\mathbf{\hat{y}}(t) = \mathbf{u}(t+\tau_f)$, according to the step ahead $\tau_f$ defined in the training phase.

\section{Detrending using ESNs}
\label{sec:ens_trend}

We now describe the main assumptions of our model and the detrending procedure to be used on a given univariate time series $y(t)$.
We consider $y(t)$ as being composed of two superimposed components of different degrees of predictability:
\begin{itemize}
\item{a trend process $x(t)$, which corresponds to the main stochastic process. This process represents the intrinsic dynamical evolution of the studied system and is predictable with high accuracy by an ESN;}
\item{a noise process $n(t)$, which is less predictable by an ESN, hence requiring a more complex model to be described.}
\end{itemize}
Under the assumption of statistical independence between $x(t)$ and $n(t)$, $y(t)$ can be separated in the sum
\begin{equation}
\label{eq:yxn}
y(t) = x(t) + n(t), \;\;\; t \in \mathbb{N}.
\end{equation}
The trend $x(t)$ is a nonstationary stochastic process of larger magnitude with respect to $n(t)$, even if there are no hard constraints on their relative scales. The noise process, instead is a zero-mean, self-similar and stationary stochastic process which can in general be correlated, and thus is characterized by a Hurst coefficient and a multifractal spectrum. 
Prototypical examples of such a process are fractional Gaussian noise and (fractional) L\'{e}vy stable processes \cite{gao2007multiscale,sanchez2014effect}.

We are interested in removing the trend process from data and in obtaining the noise component $n(t)$ in order to be able to study its fractal properties.
One way to approach this problem is to apply a filter to the measurement process and, in contrast with the common use of filters, only keep the noise part by subtracting the filtered signal from the original time series. 
A discrete-time optimal filter is a system that takes as input a measurement process $y(t)$ and outputs an estimate, $\bar{x}(t)$, of $x(t)$ at each time step $t$, such that a given error criterion (e.g., mean square error) is optimized. The simplest kind of filters are linear filters, which are widely employed in virtue of their efficiency and analytic tractability. 
However, in many situations not only the assumption of linearity is violated, but also an explicit analytical model of the signal is not available \textit{a priori}.
In these situations, it can be convenient to employ data-driven models that do not make strong assumptions on the data being processed and are capable to describe a wide range of processes. 

In this work we employ an Echo State Network (see Section \ref{sec:esn}) as a nonlinear filter in order to learn an approximation $\bar{x}(t)$ of the trend process $x(t)$, by training the system only with the measurement process $y(t)$. 
Since we are dealing with correlated noise, there is a possibility for an arbitrarily complex network to learn and predict also part of the noise process $n(t)$ and thus overfitting data. However, given our assumption of noise as a less predictable process, we constrain the neural network descriptive capability by using proper regularization techniques to prevent such overfitting.
The proposed detrending with ESN procedure, called DESN, consists of a series of steps, whose details are provided in the following.

Let us consider the pair of time series $\left\{ u_\text{data}(t), y_\text{data}(t) \right\}_{t=1}^{T}$ representing respectively the input and desired output of the network. Since in the prediction framework $y_\text{data}(t) = u_\text{data}(t+\tau_f)$, with $\tau_f$ the forecast horizon, the two time series can be constructed from a time series $\mathbf{z} = \{ z(t) \}_{t=1}^{T+\tau_f}$, representing the measurements of the observed process. 
The two time series are then split into two separate datasets: training $ \{ u_\text{tr}(t), y_\text{tr}(t) \}_{t=1}^{T_\text{tr}}$ and test set $ \{ u_\text{ts}(t), y_\text{ts}(t) \}_{t=T_\text{tr}+1}^{T}$.
The readout is trained by feeding the ESN with $u_\text{tr}(t)$ and forcing $y_\text{tr}(t)$ as teacher signal. 
At this point, the detrending procedure is applied on the remaining data of the test set. In particular, the prediction $\hat{y}_\text{ts}(t)$ is in turn utilized to detrend $y_\text{ts}(t)$, as explained below.
From now on, we assume the ESN to be already trained and then, since the training data are no longer considered, we will denote $y_\text{ts}(t)$ simply as $y(t)$.
The time series $\hat y(t)$, which denotes the values predicted by the ESN, can be expressed as:
\begin{equation}
\label{eq:predict}
\hat{y}(t) = y(t) + e_\text{pred}(t) = x(t) + n(t) + e_\text{pred}(t),\;\; t \in \mathbb{N},
\end{equation}
where $e_{\text{pred}}(t)$ is the ESN prediction error as a function of time.

The performance of a prediction model can be evaluated through the forecast accuracy, typically implemented as the normalized root mean square error \cite{de200625}, quantifying the differences between predicted and observed values.
For a given model complexity, the prediction error is related to the amount of training data and on the accuracy of the training procedure. However, even for a optimally trained model, in the presence of noise the forecast will always be subject to an error, due to (intrinsic) stochastic unpredictability of the process or insufficient complexity of the prediction model.
We refer to this source of error as \textit{intrinsic unpredictability} of the process with respect to the given model complexity and its related error function as $e_\text{intr}(t)$. 
By assuming independence between the training error $e_\text{tr}(t)$ and the intrinsic error $e_\text{intr}(t)$, we can write $e_\text{pred}(t)$ as the sum of the independent components
\begin{equation}
\label{eq:predErr}
e_{\text{pred}}(t) = e_{\text{tr}}(t) + e_{\text{intr}}(t), \;\; t \in \mathbb{N}.
\end{equation}
If the prediction model is properly trained, we can assume the training error to be negligible, i.e.,
\begin{equation}
\label{eq:perfecttraining}
e_{\text{tr}}(t) \simeq 0 \; \forall \,t \in \mathbb{N}.
\end{equation}
Our assumption in this work is that the trend process $x(t)$ of the observed signal $y(t)$ is completely predictable by an ESN model and all sources of intrinsic unpredictability are concentrated in the noise component $n(t)$. This assumption corresponds to approximating:
\begin{equation}
\label{eq:onlytrend}
\hat y(t) = \bar{x}(t) \simeq x(t) \; \forall \,t \in \mathbb{N}.
\end{equation}
When Eqs. \eqref{eq:perfecttraining} and \eqref{eq:onlytrend} hold, by inserting Eq. \eqref{eq:predErr} in \eqref{eq:predict} we obtain:
\begin{equation}
n(t) \simeq -e_\text{intr}(t) \; \forall \,t \in \mathbb{N}.
\end{equation}
In this case, the predicted time series, $\hat{y}(t)$, is a good approximation $\bar{x}(t)$ of the trend component $x(t)$ of $y(t)$. Therefore, an estimation $\bar{n}(t)$ of the true noise $n(t)$ can be obtained as:
\begin{equation}
\label{eq:noiseEval}
\bar{n}(t) \equiv y(t) - \hat{y}(t) = -e_\text{pred}(t) \simeq -e_\text{intr}(t).
\end{equation}
The time series that we analyze here contains measurements of a signal with a superimposed noise, which increases the difficulty of obtaining high reliability in short-term forecasts. For this reason, one needs to wait until the trend accumulates sufficiently before it becomes clear: considering different forecast horizons could significantly influence the result of the prediction. 
In order to mitigate the dependency of the prediction performance on the particular forecast horizon $\tau_f$, we perform multiple forecasts using an ensemble of $k$ independent ESNs, each one trained considering a different prediction step-ahead $\tau_f^{(i)}, i=1, ..., k$.
The output signals of the ensemble of predictors, elaborated on the basis of the same input data but using different forecast horizons, generate independent outcomes $\hat{y}_i(t), i=1, ..., k$, that are combined together in an average forecast, $\hat{y}(t) = 1/k \sum_{i=1}^k \hat{y}_i(t)$. This approach provides a more accurate prediction by compensating for the variance introduced by the single predictors.
Such an approach is related to the well-known frameworks of ensemble learning \cite{dietterich2000ensemble,topchy+jain+punch2005} and neural network ensembles \cite{hansen1990neural}. In the latter it has been shown experimentally that the synergy of multiple back-propagation neural networks improved learning, generalization capability, noise tolerance, and self-organization with respect to a single, yet more complex system.

\subsection{Other detrending methods}
\label{sec:detrendMethods}

In this section, we describe some existing methodologies that have been used in previous works for separating trends from the noise components in a time series \cite{bashan2008comparison}. To be consistent with our approach, we consider the following detrending procedures as MFDFA preprocessing steps.

\paragraph{Empirical Mode Decomposition}
\label{par:EMD}
Empirical Mode Decomposition (EMD) is a data-driven technique that performs a decomposition of the original signal, $y(t)$, in terms of a finite number of modes $g_i(t)$, called Intrinsic Mode Functions (IMF), and a residual component.
IMFs are derived directly from data, without any prior assumption about their model.
EMD \cite{flandrin2004empirical} can be used to extrapolate a trend in data by considering the residual given by: $\bar{x}(t)=y(t)-\sum_{i=1}^n g_i(t)$. The residue is hence subtracted from the original time series in order to remove the global trend and obtain an estimate of the noise. Generally, as shown in \citet{wu2007trend}, also a number of IMFs are selected along the residual in order to better approximate the trend. This is especially needed where the trend is composed by periodicities, which cannot be approximated by a single residual. 
The EMD procedure has also been applied as a local detrending method in the windows computed with DFA, in place of the conventional polynomial fitting \cite{qian2011modified}.

\paragraph{Fourier-Detrended Fluctuation Analysis}
\label{par:FD}
The Fourier-Detrended Fluctuation Analysis (FDFA) is a tool used for identifying trends characterized by frequencies with a significant power \cite{movahed2008fractal}. The method targets the first few coefficients (those having larger amplitude or real part) of a Fourier expansion and thus it can be considered as a simple high-pass filter \cite{chianca2005fourier}.
We use a slightly different approach here, which consists in cutting the spectral components with higher amplitude, rather than exclusively focusing on those having lower frequencies -- as originally proposed in \cite{chianca2005fourier}. In this way, the definition of trends is relaxed in order to consider all larger amplitude periodicities, independently of their variation speed.
Specifically, we first apply the discrete fast Fourier transform to the data records, then we sort the spectral components according to a decreasing order of their amplitude. Successively, we truncate the first $\tau_{\text{freq}}$ coefficients of the Fourier expansion.
Finally, we apply the inverse Fourier transform to the truncated series. After this last step, \textit{border effects} may appear at the opposite ends of the time series. These distortions are eliminated by cropping a portion of the initial and last part of the series.

\paragraph{Smoothing}
\label{par:SM}
Smoothing methods operate in the time domain and basically implement low-pass filters. High frequency are attenuated on the base of the specific properties of the adopted smoothing method. We consider four different smoothing procedures, which depend on a parameter $\sigma$, representing the span of the smoothing procedure:
\begin{itemize}
	\item Algorithm 1: a low-pass filter with coefficients equal to the reciprocal of the span (moving average);
	\item Algorithm 2: local regression using weighted linear least squares and a 1st degree polynomial model;
	\item Algorithm 3: local regression using weighted linear least squares and a 2nd degree polynomial model;
	\item Algorithm 4: a generalized moving average with filter coefficients determined by an unweighted linear least-squares regression and a polynomial model of specified degree $p$.
\end{itemize}


\section{Experimental results}
\label{sec:exp}

In this section, we evaluate the performance of DESN, the proposed detrending method based on ESN. We compare the results with those obtained using the detrending methods introduced in Section \ref{sec:detrendMethods}, namely Empirical Mode Decomposition (EMD), Fourier-Detrended Fluctuation Analysis (FDFA), and different Smoothing (SM) techniques.
In order to demonstrate the effectiveness of the proposed technique, we consider several synthetic time series having a self-similar noise component with known characteristics.
We also test the methods on a real-world dataset, the sunspot time series, described in Section \ref{sec:sunspot}.
These latter data have already been studied in the (multi)fractal analysis context -- see, for example, \cite{hu2009multifractal,drozdz2015detecting} and references therein. The datasets taken into account and a MATLAB code for reproducing all experiments presented in this paper are publicly available\footnote{\texttt{https://bitbucket.org/slackericida/desn\_v1/overview}}.

\subsection{Synthetic time series}
\label{sec:synth}

As described above, the synthetic time series are of the form $y(t) = x(t) + n(t)$, with $x(t)$ the trend and $n(t)$ the noise component. We use the four aforementioned detrending methods for computing an estimation of $x(t)$, namely $\bar{x}(t)$, and we evaluate the accuracy of each method by analyzing the LTC and multifractal properties of the estimated noise, $\bar{n}(t)= y(t) - \bar{x}(t)$.
The accuracy of each method is evaluated by comparing the coefficients obtained with MFDFA (see Section \ref{sec:mfdfa}) on the estimated noise $\bar{n}(t)$ with respect to the ground-truth $n(t)$. For all the synthetic series and methods, the MFDFA procedure has been executed on scales ranging from 16 to 1024 data points and with a second-order local polynomial detrending. The parameter $q$ ranges from -5 to +5.

We consider seven time series \texttt{Y1}, \dots,\texttt{Y7}, which are obtained by combining a trend selected from one of the five different time series \texttt{X1}, \dots,\texttt{X5} with a noise selected from one of the three different time series \texttt{n1}, \texttt{n2}, and \texttt{n3}.
Signals used as trend are described by the functions shown in Table \ref{tab:trendDesc}.
For the trend signals, \texttt{X1},\texttt{X2},\texttt{X4}, and \texttt{X5}, we report the interval from which the values of the domain variable $x$ are extracted.
In Table \ref{tab:noiseDesc} are summarized the average properties of the synthetic noise components. 
We use two different sets of ten fGn processes generated by setting $H$ respectively to 0.7 and 0.3, and a deterministic binomial multifractal cascade \cite{PhysRevE.74.016103} with multiplicative factor equal to $0.60708$.
For the noise $\texttt{n3}$, we also consider the spectrum asymmetry
\begin{equation}
\label{eq:asymmetry}
\Theta = \frac{\Delta \alpha_\text{L} - \Delta \alpha_{R}}{\Delta \alpha_\text{L} + \Delta \alpha_{R}},
\end{equation}
where $\Delta \alpha_\text{L}$ and $\Delta \alpha_\text{R}$ are the width of the left and right part of the support of $D(\alpha)$ (\ref{eq:mutifractal_spectrum}), respectively.
A negative value for $\Theta$ denotes a right-sided spectrum, highlighting a stronger multifractality on smaller fluctuations, while the contrary holds in the case of a positive value.
All time series have been normalized by calculating the z-score; the amplitudes of signal and noise series are multiplied by a suitable scalar value, in order to obtain a signal-to-noise ratio of 16.
\bgroup
\def\arraystretch{1.5} 
\setlength\tabcolsep{1em} 
\begin{center}
\begin{table}[ht]\small
\centering
\caption{Description of the functions used as trend within the synthetic signals. The term $\nu_\text{max}$ refers to the Nyquist frequency $f_\text{s}/2$, where $f_\text{s}$ is the sampling rate, and the terms $\mathcal{U}(x_\text{min}, x_\text{max})$ and $\mathcal{N}(\mu_x,\sigma_x)$ are respectively the uniform and normal distributions.}
\vspace{0.2cm}
\begin{tabular}{ll}
\hline
\textbf{ID} & \textbf{Description} \\
\hline
\texttt{X1} & $\sin(t)$. \\
\texttt{X2} & $\sum_{i=1}^{10} A_i \sin(2\pi \nu_i t), \;\;\; \left\{ \nu_i = \mathcal{U}(0,10^{-5}\nu_\text{max}) \right\}, \left\{A_i = \mathcal{N}(1,1)\right\}$. \\
\texttt{X3} & $s\|s\|s\|...$, with $s$ the first 100 digits of $\pi$. \\
\texttt{X4} & $\sum_{i=1}^{10} A_i \sin(2\pi \nu_i t), \;\;\; \left\{ \nu_i = \mathcal{U}(0,0.5\nu_\text{max}) \right\}, \left\{A_i = \mathcal{N}(1,1)\right\}$. \\
\texttt{X5} & $\sin(t)/t^2$. \\
\hline
\end{tabular}
\label{tab:trendDesc}
\end{table}
\end{center}
\egroup

\bgroup
\def\arraystretch{1.5} 
\setlength\tabcolsep{1em} 
\begin{center}
\begin{table}[ht]\small
\centering
\caption{Characteristics of the synthetic noise processes. The Hurst exponent and MFW of \texttt{n1} and \texttt{n2} are the outcome of MFDFA averaged over ten independent realizations of the process.}
\vspace{0.2cm}
\begin{tabular}{lllll}
\hline
\textbf{ID} & \textbf{Description} & \textbf{Length} & \textbf{avg. Hurst} & \textbf{avg. MFW} ($\Theta$)\\
\hline
\texttt{n1} & fGn & 150000 & 0.695 & 0.022 \\
\texttt{n2} & fGn & 150000 & 0.303 & 0.032 \\
\texttt{n3} & Binomial cascade & 131072 & 0.883 & 1.192 (0.048) \\
\hline
\end{tabular}
\label{tab:noiseDesc}
\end{table}
\end{center}
\egroup

Overall, we performed seven different tests. In Table \ref{tab:testConf}, we report the time series under consideration and the values used for configuring each detrending procedure. Note that the length of the $i$-th time series \texttt{Yi} is given by the length of the noise component, which is reported in Table \ref{tab:noiseDesc}. For DESN, we consider an additional time series for training the network (referred as $y_\text{tr}(t)$ in Section \ref{sec:ens_trend}), whose length is half of \texttt{Yi}'s length.
\bgroup
\def\arraystretch{1.5} 
\setlength\tabcolsep{0.5em} 
\begin{table*}[ht!]\small
\centering
\caption{Time series and configuration of the different detrending procedures used in each test. For DESN, we report the values of the size of the reservoir ($N_r$), the spectral radius ($\rho$), the regularization coefficient ($\lambda$), and the number $k$ of forecast models. For FDFA, we report the thresholds $\tau_{\text{freq}}$ and $\tau_{\text{time}}$ used for determining the amount of coefficients to be truncated in both frequency and time domain. For SM, we report the span of the moving average $\sigma$ and the identifier of the adopted algorithm. Finally, for EMD we report the number of the last $s$ IMFs which are used for defining the trend.}
\vspace{0.2cm}
\hspace*{-0.8cm}
\begin{tabular}{llllll}
\hline
\textbf{Data} & \centering\textbf{DESN} & \textbf{FDFA} & \textbf{SM} & \textbf{EMD}\\
\hline
\texttt{Y1} = \texttt{X1} + \texttt{n1} & \parbox[c][1.2cm]{2.8cm}{$N_r = 500$, $\rho = 0.99$,\\ $\lambda = 0.1$, $k = 30$} & $\tau_{\text{freq}} = 150$, $\tau_{\text{time}} = 950$ & $\sigma = 50$, algo: 2 & $s=13$\\
\texttt{Y2} = \texttt{X2} + \texttt{n1} & \parbox[c][1.2cm]{2.8cm}{$N_r = 200$, $\rho = 0.4$,\\ $\lambda = 0.1$, $k = 20$} & $\tau_{\text{freq}} = 60$, $\tau_{\text{time}} = 1$ & $\sigma = 1800$, algo: 3 & $s=5$\\
\texttt{Y3} = \texttt{X3} + \texttt{n1} & \parbox[c][1.2cm]{2.8cm}{$N_r = 500$, $\rho = 0.99$,\\ $\lambda = 0.1$, $k = 20$} & $\tau_{\text{freq}} = 115$, $\tau_{\text{time}} = 50$ & $\sigma = 20$, algo: 4 & $s=19$\\
\texttt{Y4} = \texttt{X4} + \texttt{n1} & \parbox[c][1.2cm]{2.8cm}{$N_r = 400$, $\rho = 0.99$,\\ $\lambda = 0.1$, $k = 10$} & $\tau_{\text{freq}} = 400$, $\tau_{\text{time}} = 3000$ & $\sigma = 10$, algo: 1 & $s=17$\\
\texttt{Y5} = \texttt{X5} + \texttt{n1} & \parbox[c][1.2cm]{2.8cm}{$N_r = 100$, $\rho = 0.99$,\\ $\lambda = 0.05$, $k = 30$} & $\tau_{\text{freq}} = 4000$, $\tau_{\text{time}} = 250$ & $\sigma = 1000$, algo: 1 & $s=8$\\
\texttt{Y6} = \texttt{X1} + \texttt{n2} & \parbox[c][1.2cm]{2.8cm}{$N_r = 500$, $\rho = 0.99$,\\ $\lambda = 0.1$, $k = 30$} & $\tau_{\text{freq}} = 400$, $\tau_{\text{time}} = 2000$ & $\sigma = 50$, algo: 2 & $s=17$\\
\texttt{Y7} = \texttt{X1} + \texttt{n3} & \parbox[c][1.2cm]{2.8cm}{$N_r = 500$, $\rho = 0.99$,\\ $\lambda = 0.05$, $k = 20$} & $\tau_{\text{freq}} = 250$, $\tau_{\text{time}} = 2000$ & $\sigma = 60$, algo: 4 & $s=24$ \\
\hline
\end{tabular}
\label{tab:testConf}
\end{table*}
\egroup

Results are obtained by averaging ten independent realizations of the tests.
The sources of randomicity for each test are the different realizations of the noise process -- for $\texttt{n1}$ and $\texttt{n2}$ -- and the different executions of the DESN procedure -- ESN input and reservoir weights. 
We used a grid search to tune the (hyper-)parameters of the different methods in their respective spaces.
For each detrending method, we considered a different sets of bounds and search resolutions of the respective parameter space and a specific loss function for guiding the optimization.
The error measurement that we used is the normalized root mean squared error (NRMSE) function, which is defined as follows:
\begin{equation}
\label{eq:nrmse}
\textrm{NRMSE} = \sqrt{\frac{\langle \lVert \mathbf{y} - \mathbf{d} \rVert^2 \rangle}{\langle \lVert \mathbf{y} - \langle\mathbf{d}\rangle \rVert^2 \rangle}},
\end{equation}
being $\mathbf{y}$ the ESN output (\ref{eq:esn_output_update}) and $\mathbf{d}$ the desired one.

\paragraph{Parameter settings of detrending methods}

For DESN, the parameters that we considered are the size $N_r$ of the reservoir, searched in $[100, 500]$ with resolution 100; the spectral radius $\rho$ searched in the set $\{0.4, 0.518, 0.636, 0.754,  0.872, 0.99\}$; 
the regularization coefficient $\lambda$ used in the linear regression for the training of the readout is searched in $[0.05, 0.3]$ with step size $0.05$; the number $k$ of forecast models used is searched in $[10, 30]$ with step size 10. 
As discussed in Section \ref{sec:ens_trend}, we used a different forecast step for training each of the $k$ ESNs of the ensemble. In particular, the forecast step of the $i$-th predictor model is $m_k = 10 \cdot i$.
The adopted loss function is the average error computed on $y$ and forecast $\hat y_i$ of the $i$-th prediction model, that is, $1/k\sum_{i=1}^k \mathrm{NRMSE}(\hat y_i, y)$.

For the SM procedure, we tuned the span of the moving average $\sigma$ in $[10, 2000]$ with step size 10. For guiding the hyper-parameter optimization, we used a loss function which minimizes the error and maximizes the span, defined as: $f_{\text{SM}} = \eta_{\text{SM}} \cdot \mathrm{Err} + (1 - \eta_{\text{SM}}) 1/\sigma$, where $\mathrm{Err}$ is the error evaluated as Eq. \eqref{eq:nrmse} and $\eta_{\text{SM}} \in [0,1]$ is a weight parameter that was set to $0.1$ in every test. Note that for $\eta_{\text{SM}} = 0$ the error component is neglected, then the resulting span is maximized covering the whole time series; this generates a smooth function which assumes in every point the mean value of the original signal.
On the other hand, by setting $\eta_{\text{SM}} = 1$, only the error is minimized and the span assumes its minimum value $\sigma = 2$, which generally produces an insufficient smoothing of the signal.
We evaluated the performances using all the four algorithms described in Section \ref{par:SM} and we reported here the one which achieved the best results. The polynomial degree $p$ in the algorithm 4 was set to 15 in every test.

For setting the optimal values of the parameter $\tau_{\text{freq}}$ in the FDFA procedure, after having ordered the Fourier coefficients by their amplitude (from larger to smaller), by visual inspection we first identify the ``elbow'' in the sequence, which is its inflection point, which determines the frequencies to be truncated (i.e., these having very high power).
Once the inverse Fourier transform is performed, some cropping on the boundaries of the time series is necessary to attenuate boundary effects caused by the alteration of the spectrum. 

Finally, in the EMD approach we used the standard setup of the stop criterion for retrieving the IMFs, as described in \cite{Huang2317}. The sum of the last $s$ IMFs represents the trend and the number $s$ is optimized by minimizing the following loss function: $f_{\text{EMD}} = \eta_{\text{EMD}} \mathrm{Err} + (1 - \eta_{\text{EMD}}) s/S$, where $S$ represents the total number of IMFs identified relative to each signal -- usually between 15 and 20 components. Also in this case, $\mathrm{Err}$ is the error evaluated with Eq. \eqref{eq:nrmse} and $\eta_{\text{EMD}} \in [0,1]$ is a weight parameter. Note that for $\eta_{\text{EMD}} = 0$ the error component is neglected and $s$ assumes the minimum value 1, i.e., only the last IMF is selected for approximating the trend. On the other hand, when $\eta_{\text{EMD}} = 1$ the error is minimized, but all the $s$ IMFs are selected for representing the trend, which then coincides with the original signal.
We set $\eta_{\text{EMD}} = 0.1$ when we tested the synthetic signals \texttt{Y3}, \texttt{Y4}, \texttt{Y6}, and \texttt{Y7}, while in the processing of the remaining signals (including the sunspot time series) we set $\eta_{\text{EMD}} = 0.5$.

\paragraph{Discussion of results}

\begin{figure*}[ht!]
\centering
\subfigure[\texttt{Y1}]{
\includegraphics[width=0.48\textwidth]{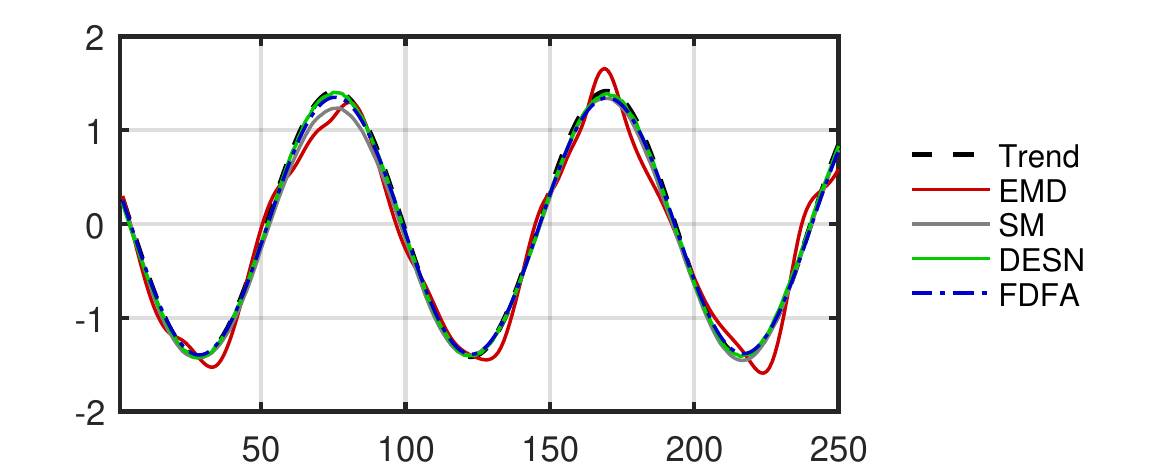}}
~
\subfigure[\texttt{Y2}]{
\includegraphics[width=0.48\textwidth]{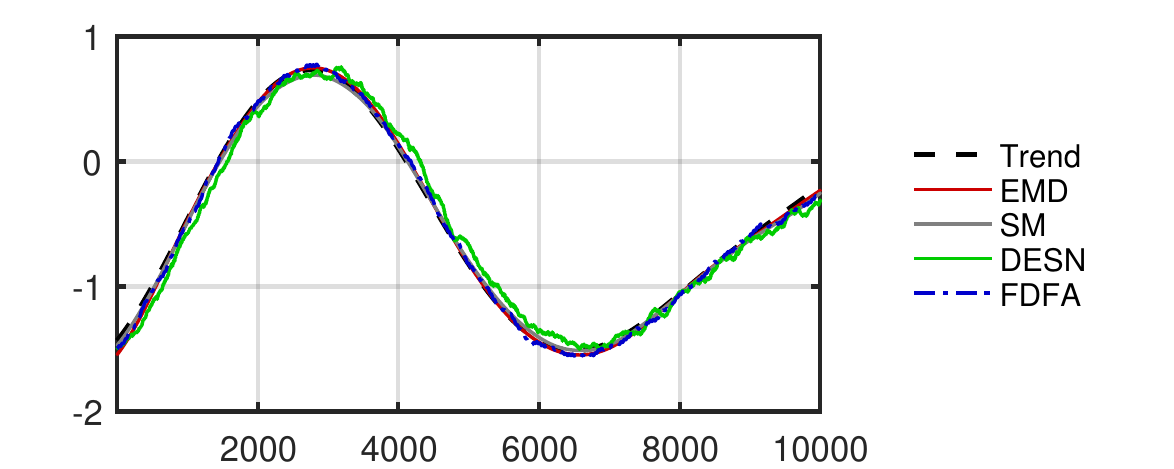}}

\subfigure[\texttt{Y3}]{
\includegraphics[width=0.48\textwidth]{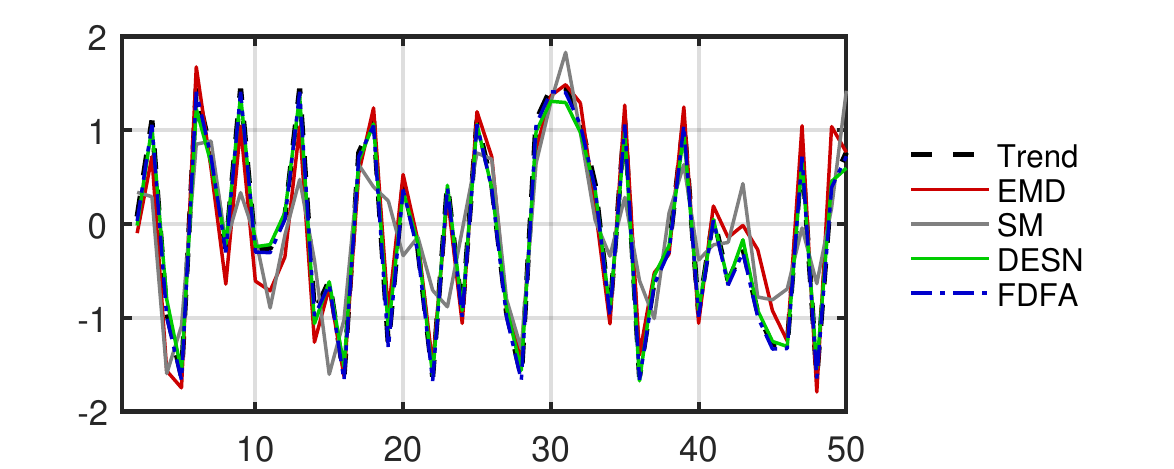}}
~
\subfigure[\texttt{Y4}]{
\includegraphics[width=0.48\textwidth]{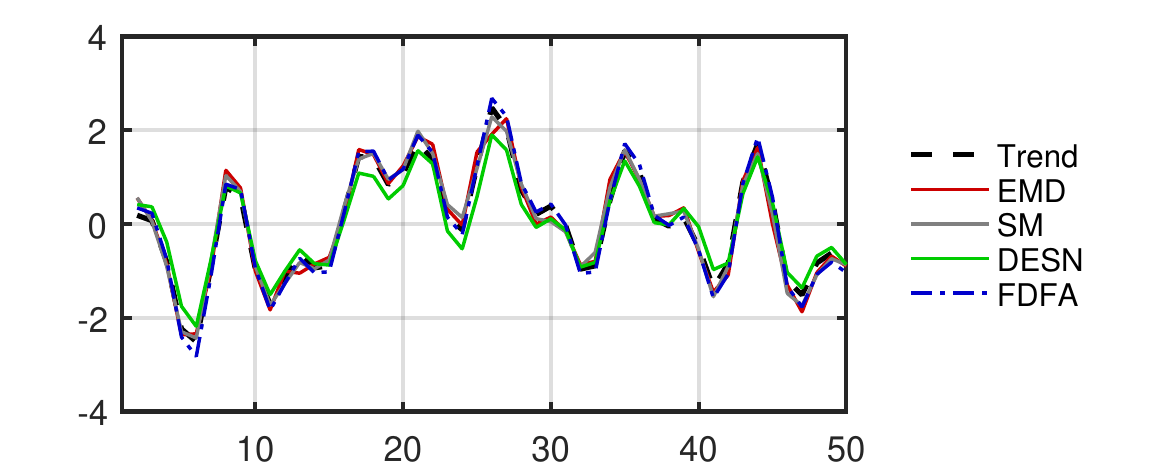}}

\subfigure[\texttt{Y5}]{
\includegraphics[width=0.48\textwidth]{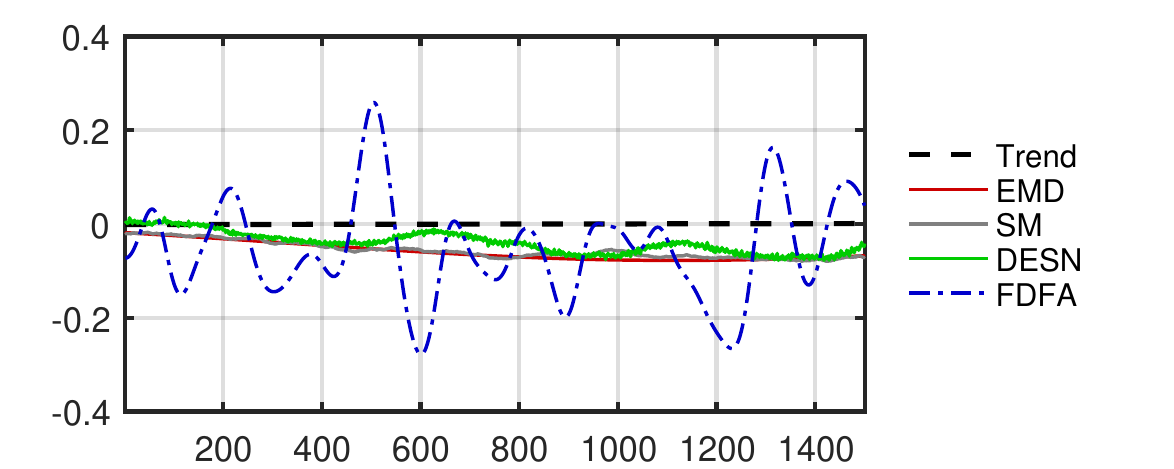}}
~
\subfigure[\texttt{Y6}]{
\includegraphics[width=0.48\textwidth]{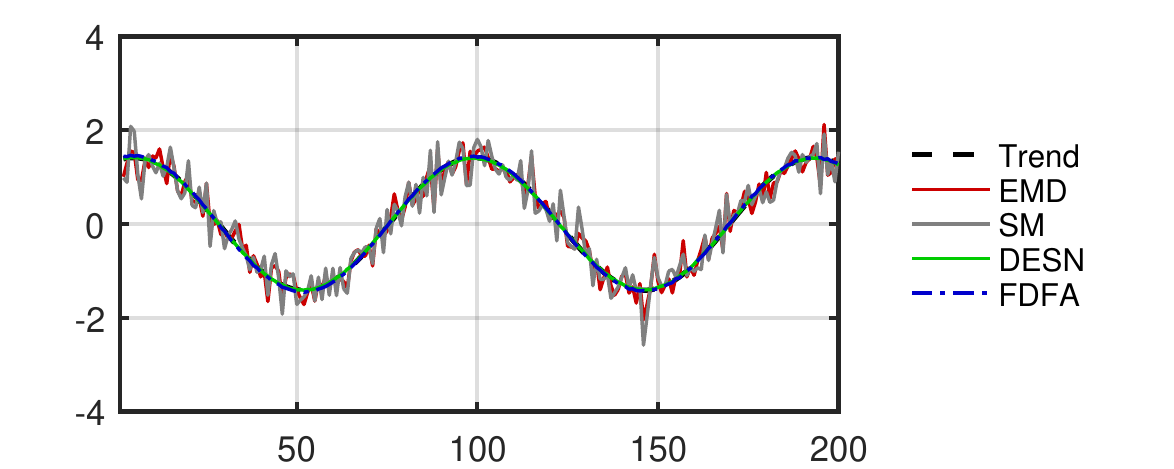}}

\subfigure[\texttt{Y7}]{
\includegraphics[width=0.48\textwidth]{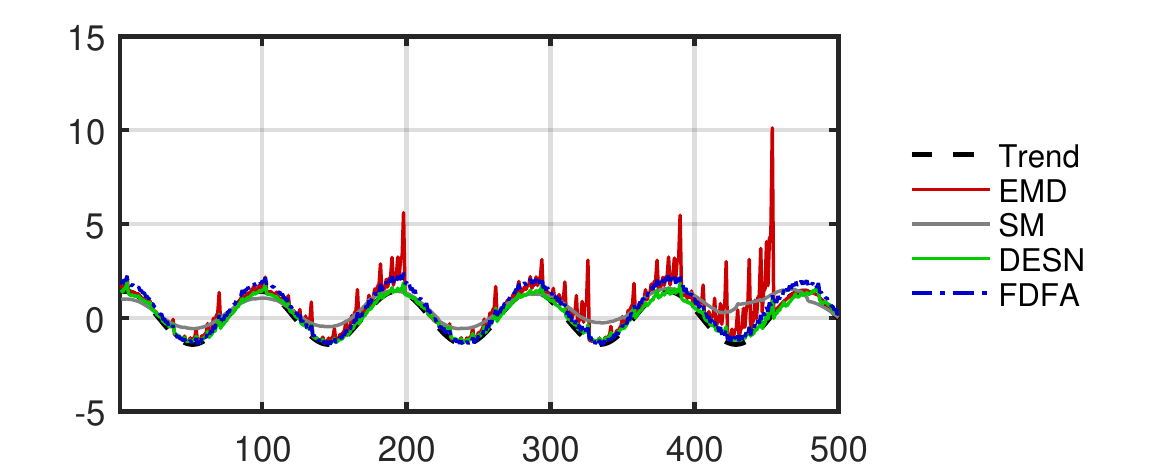}}

\caption{Colors online. Trends identified on the different signals. The function depicted with black dashed lines represents the trend of the original time series. The colored lines represent the trends identified using DESN, EMD, FDFA, and SM. For clarity of representation only small portions of the time series are shown.}
\label{fig:trends}
\end{figure*}

\begin{figure*}[ht!]
\centering
\includegraphics[keepaspectratio,scale=0.7]{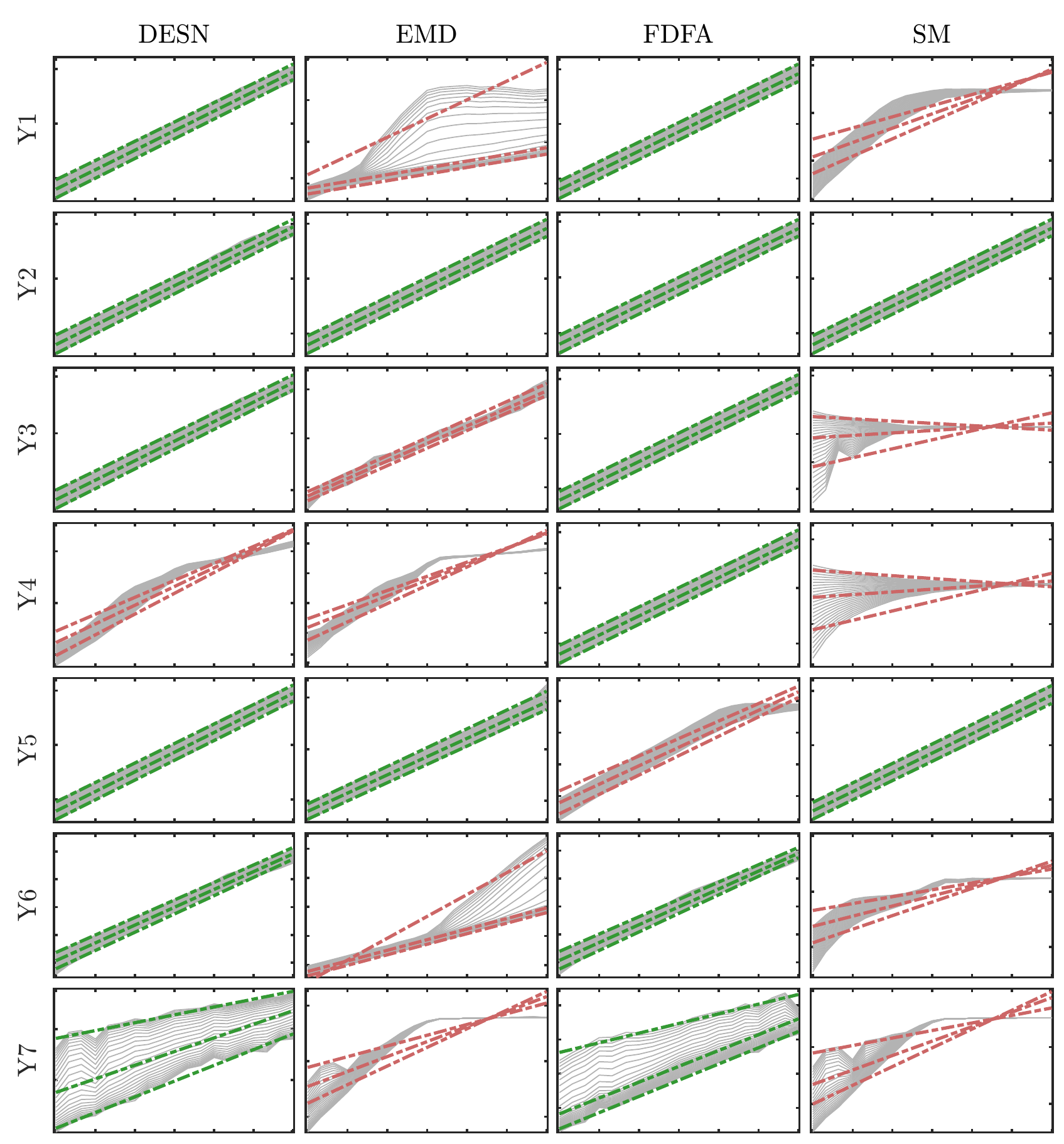}
\caption{Colors online. Scaling of fluctuation functions related to the detrended time series. Only one instance of each test in Table \ref{tab:results} is reported here. The least-square linear fittings are highlighted in green when they correspond to a correct scaling function and in red otherwise.}
\label{fig:scalingSeries}
\end{figure*}

In Fig. \ref{fig:trends}, we plot a short sample of each time series with superimposed the trends identified by the different detrending procedures.
The details of the results are reported in Table \ref{tab:results}, where we show the resulting Hurst coefficient and multifractal spectrum width (MFW) for each time series, with their corresponding standard deviations. 
In Fig. \ref{fig:scalingSeries} we graphically represent the quality of the scaling of the fluctuation function for the estimated noise components. The linear fittings of the scaling functions are highlighted in green when the considered detrending method (column) has preserved a correct scaling behavior on the selected time series (row), while we used red dashed linear fittings to denote an incorrect scaling or significantly altered Hurst/MFW coefficients with respect to the ground truth.

As shown in Table \ref{tab:results}, the four methods perform differently on each time series. With the EMD and SM methods, and considering the parameter optimization criteria presented in Section \ref{sec:detrendMethods}, we could not obtain a correct scaling for most of the tested time series.
The first five time series, $\texttt{Y1-5}$, are composed by a signal (trend) with a superimposed persistent noise with $H=0.7$, according to Table \ref{tab:testConf}. 

In $\texttt{Y1}$, the trend is a single sinusoid, which is the simplest periodic function and it is easily separable from noise, which is much more complex from a prediction perspective. As expected, the Hurst exponent is estimated with a good precision by DESN. FDFA obtains a similar accuracy, since in this case the trend can be easily isolated, it being described by a single high-amplitude frequency in the Fourier domain. In fact, as described in Section \ref{sec:detrendMethods}, FDFA operates by eliminating the frequencies with largest amplitudes, so its maximum efficiency is reached when trends consists of few isolated dominating frequencies.
On the other hand, in time series where trend periodicities are spread over a large portion of the spectrum or are too entwined with the noise frequencies, FDFA tends to fail. In fact, by cutting a significant amount of frequencies, FDFA tends to corrupt the spectrum of noise and hence its scaling properties. It is important to point out that the original FDFA method proposed in \cite{chianca2005fourier} works only as a low-pass filter without taking into account amplitudes, so its limitation is even more evident in these particular cases.
The SM and EMD procedures do not perform well on identifying the trend in $\texttt{Y1}$. While this is a common issue with EMD applied to sinusoidal signals \cite{wu2007trend}, with SM we can observe in the example of Fig. \ref{fig:scalingSeries} a crossover that breaks the global scaling. This crossover is given by the smoothing algorithm acting only at a scale determined by its span parameter. 

Despite the apparent increasing difficulty of the detrending task on the second time series $\texttt{Y2}$, whose trend is a linear combination of low-frequency sinusoids with different amplitudes, all methods perform equally well. However, by comparing the trend functions in Table \ref{tab:trendDesc}, it is important to notice that the frequency of the sinusoid function in $\texttt{X1}$ is significantly higher than the maximum value of the frequencies characterizing the trend $\texttt{X2}$. 
In this case, in fact, the variation of the trend signal is sufficiently slow to be isolated properly by EMD and SM, which behave in this case as low-pass filters.

On the third series $\texttt{Y3}$, the results are similar to what observed in the first test. In fact, the trend signal is a periodic series obtained by repeatedly concatenating the first 100 digits of $\pi$. Therefore, the trend is characterized by a broad spectrum with fast frequencies, and thus EMD and SM are once again unable to perform the required task.
In fact, even if from Fig. \ref{fig:scalingSeries} we can observe the log-log scaling of EMD to be approximately linear, the obtained Hurst coefficient is 0.366, which differs significantly from the true value of 0.695 and incorrectly denoting an antipersistent behavior. This means that the fractal properties of noise have been considerably altered by the EMD detrending procedure and the result is not to be considered correct. 

The trend in $\texttt{Y4}$ is a more complex version of $\texttt{Y2}$, since the signal $\texttt{X4}$ is characterized also by high frequencies, it being composed by a linear combination of 10 sine waves with frequencies chosen randomly in a broad interval.
In this case, only FDFA succeeds in detrending the series correctly, since the spectrum of the trend consists of isolated high-amplitude frequencies.
In fact, as explained above, the FDFA procedure implemented in this work filters the spectral components with greater amplitudes, regardless of their frequency, thus making the filtering method independent of the variation speed of the signal. EMD and SM, instead, are designed with the underlying assumption that trends are characterized by low frequencies (slow variation) and hence they are unable to filter rapidly-varying trends correctly.
DESN, on the other hand, does not perform any explicit assumption regarding the form of the trend. 
In this case, however, the resulting signal is much harder to predict since its periodicity is much longer than the network's memory can account for. In particular, it has been shown that ESNs are unable to learn functions composed of even two superimposed oscillators with incommensurable frequencies \cite{Jaeger78}, because of the aperiodicity of the compound signal. Such a signal, in fact, would require the simultaneous coexistence of two stable and uncoupled oscillating modes in the network's dynamics, a configuration that is very difficult to attain in practice.

The time series $\texttt{Y5}$ is instead a classic example where the FDFA method fails. In this case, the trend signal does not consist of isolated frequencies, but it is described by a continuous distribution of frequencies in the spectrum, most of them characterized by a small amplitude.
Hence, the filtering performed by FDFA alters the signal and this results in a crossover at larger scales, as we observe in Fig. \ref{fig:scalingSeries}. All the other methods, instead, perform well on this time series, given the regular behavior of its trend signal in the time domain and the prevalence of low frequencies in the Fourier domain.

The time series $\texttt{Y6}$ is composed by the trend $\texttt{X1}$ with the addition of antipersistent noise. Analogously as what observed for $\texttt{Y1}$, only DESN and FDFA succeed in correctly identifying the trend on such a time series. 
So far, in every test the estimation of \texttt{n1} and \texttt{n2} resulted to be monofractal, as confirmed by the estimated MFWs shown in Table \ref{tab:results}. The only exception is in the outcome given by EMD on $\texttt{Y5}$, where we detect on $\bar{n}(t)$ the presence of spurious multifractal scaling, which is not present in the ground-truth signal \texttt{n1}.

The time series $\texttt{Y7}$ is the only series characterized by a multifractal scaling. As shown in the results, in this case only DESN and FDFA produce a correct scaling function, even if the precision of the estimation is not optimal, probably because of the higher complexity of such a time series.
The calculated Hurst coefficient is (slightly) overestimated by DESN and underestimated by FDFA. The principal difference in performance between these two approaches lies in the estimated multifractal spectrum width. In fact, in this case the estimate obtained with DESN is significantly closer to the ground truth, while FDFA considerably underestimates its value, thus suggesting a process with far less multifractal properties. Moreover, we can observe that both methods overestimate the asymmetry with a bias on the left-hand side of the spectrum.
In the case of DESN, this can be explained by considering that the right-hand side of the spectrum corresponds to the smaller fluctuations, which are more easily affected by the ESN prediction error.
\bgroup
\def\arraystretch{1.5} 
\setlength\tabcolsep{1em} 
\begin{table*}[th]\small
\centering
\caption{Average values and standard deviations (where applicable) of Hurst exponent and width of the multifractal spectrum (MFW) of the noise estimated on each time series along with the ground truth (GT) value evaluated on the original noise. The asymmetry $\Theta$ of the multifractal spectrum (Eq. \eqref{eq:asymmetry}) of the series \texttt{Y7} is reported in brackets. The standard deviation is not defined for the results of FDFA on series \texttt{Y7}, since the values are deterministic. The cases in which the detrending method did not succeed in preserving the noise self-similarity are denoted with ``n.s.''.}
\vspace{0.3cm}
\begin{tabular}{cllllll}
\cline{2-7}
\multirow{8}{*}{\centering\rotatebox{90}{\hspace{-1.5em}\textbf{Hurst}}} & \textbf{ID} & \textbf{GT} & \textbf{DESN} & \textbf{FDFA} & \textbf{SM} & \textbf{EMD} \\
\cline{2-7}
& \texttt{Y1} & 0.695 & 0.713 $\pm$ 0.007 & 0.705 $\pm$ 0.007 & n.s. & n.s. \\ 
& \texttt{Y2} & 0.695 & 0.719 $\pm$ 0.007 & 0.690 $\pm$ 0.004 & 0.706 $\pm$ 0.005 & 0.701 $\pm$ 0.006 \\ 
& \texttt{Y3} & 0.695 & 0.691 $\pm$ 0.006 & 0.702 $\pm$ 0.006 & n.s. & 0.366 $\pm$ 0.004\\ 
& \texttt{Y4} & 0.695 & n.s. & 0.687 $\pm$ 0.002 & n.s. & n.s.\\ 
& \texttt{Y5} & 0.695 & 0.718 $\pm$ 0.006 & n.s. & 0.711 $\pm$ 0.006 & 0.711 $\pm$ 0.007\\ 
& \texttt{Y6} & 0.303 & 0.318 $\pm$ 0.003 & 0.314 $\pm$ 0.002 & n.s. & n.s.\\
& \texttt{Y7} & 0.883 & 1.021 $\pm$ 0.003 & 0.793  & n.s. & n.s.\\
\cline{2-7}
\multirow{8}{*}{\centering\rotatebox{90}{\hspace{0.5em}\textbf{MFW} ($\Theta$)}} & \texttt{Y1} & 0.022 & 0.027 $\pm$ 0.012 & 0.026 $\pm$ 0.006 & n.s. & n.s. \\ 
& \texttt{Y2} & 0.022 & 0.032 $\pm$ 0.014 & 0.034 $\pm$ 0.013 & 0.028 $\pm$ 0.011 & 0.023 $\pm$ 0.009\\ 
& \texttt{Y3} & 0.022 & 0.029 $\pm$ 0.008 & 0.024 $\pm$ 0.006 & n.s. & 0.023 $\pm$ 0.005\\ 
& \texttt{Y4} & 0.022 & n.s. & 0.037 $\pm$ 0.010 & n.s. & n.s.\\ 
& \texttt{Y5} & 0.022 & 0.019 $\pm$ 0.007 & n.s. & 0.018 $\pm$ 0.005 & 0.102 $\pm$ 0.041\\ 
& \texttt{Y6} & 0.032 & 0.040 $\pm$ 0.008 & 0.043 $\pm$ 0.002 & n.s. & n.s. \\
& \texttt{Y7} & \parbox[t][0.8cm]{1cm}{1.192 \\ (0.048)} & \parbox[t][0.8cm]{2.2cm}{1.116 $\pm$ 0.046 \\ (0.397 $\pm$ 0.060)} & \parbox[t][0.8cm]{2.2cm}{0.593 (0.849)} & n.s. & n.s. \\
\cline{2-7}
\end{tabular}
\label{tab:results}
\end{table*}
\egroup

\subsection{Sunspot data}
\label{sec:sunspot}

In this section, we consider the time series relative to the number of daily sunspots \cite{sunspot}. The dataset contains more than 70000 records and is characterized by a trend given by the well-known 11-year cycle of the sun. Such a dataset has been already used by other authors in the field of (multi)fractal time series analysis (see, e.g., \cite{drozdz2015detecting,hu2009multifractal}).
For all the methods taken into account here, the MFDFA procedure has been executed on the detrended series with scale parameter ranging from 16 to 1024 data points, first-order local polynomial detrending, and parameter $q$ ranging from -5 to +5.

For this test, we configured FDFA with $\tau_{\text{freq}} = 150$ and $\tau_{\text{time}} = 500$. In the EMD case, we set the weight parameter $\eta_{\text{EMD}} = 0.5$ in the cost function. For SM, we set the span $\sigma = 1000$, the weight parameter $\eta_{\text{SM}} = 0.1$, and we used algorithm 2. For DESN, we set the reservoir size $N_r = 500$, the regularization coefficient $\lambda = 0.05$, and the spectral radius $\rho = 0.99$.
For DESN, we compared two settings with different numbers $k$ of forecast models, namely $k=10$ and $k=30$, which produced slightly different, yet qualitatively comparable results. 
Since there is no known ground truth for the sunspot time series, in this section we compare our results with the properties reported in other works \cite{drozdz2015detecting,hu2009multifractal}.

\begin{figure*}[ht!]
\centering
\includegraphics[width=\textwidth]{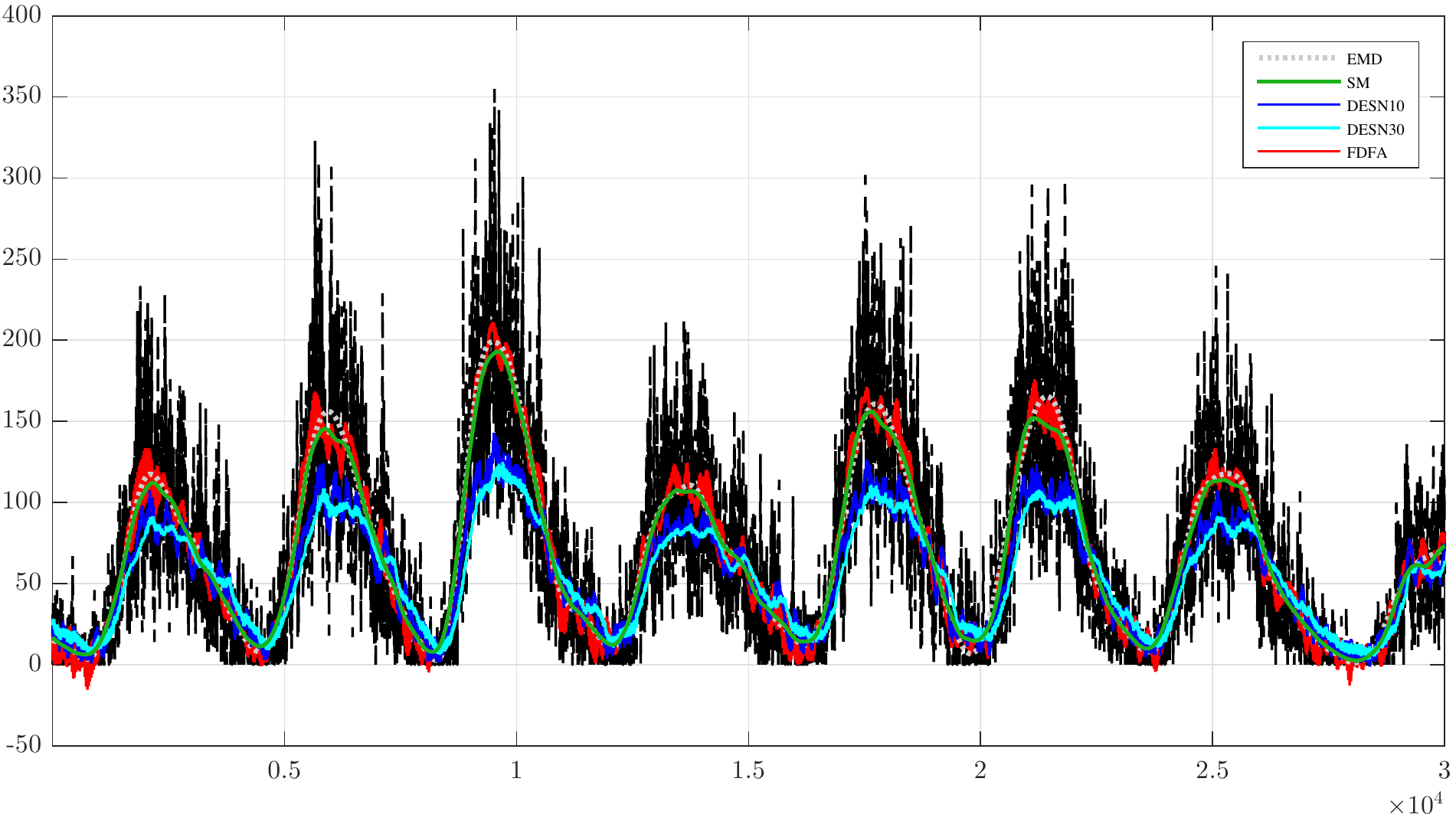}
\caption{Colors online. Trends identified on the sunspot time series. The function depicted with black dashed lines represent the original time series. The colored lines represent the trends identified using EMD, FDFA, SM, and DESN with $k=10$ and $k=30$.}
\label{fig:trendSun}
\end{figure*}

\begin{figure*}[ht!]
\centering
\includegraphics[width=\textwidth]{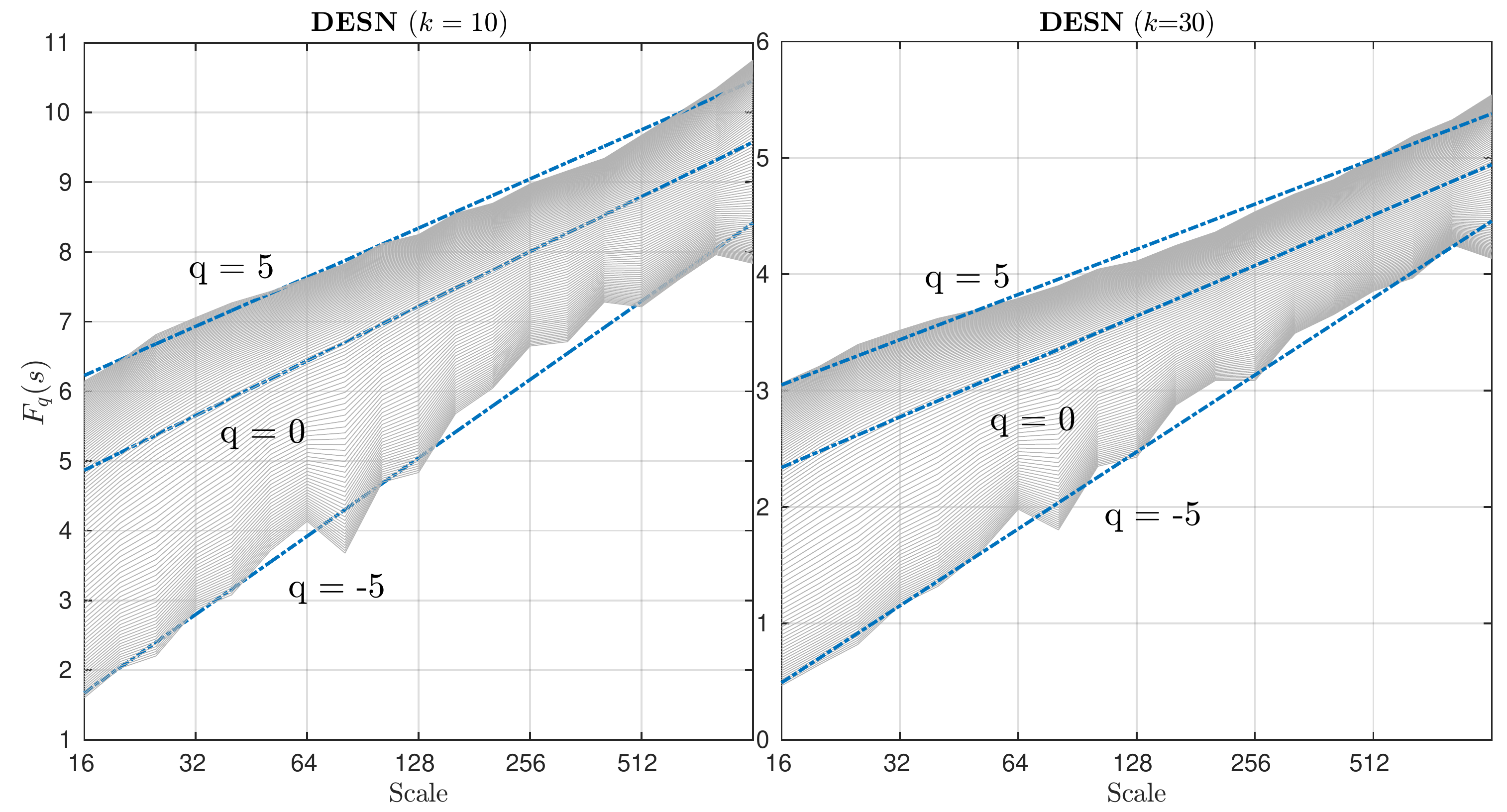}
\caption{Scaling properties of the (detrended) sunspot time series obtained with DESN for two settings of the ensemble parameter $k$.}
\label{fig:scalingSun}
\end{figure*}

In Table~\ref{tab:resultsSunspot}, we show the values of the Hurst coefficient and the width of the multifractal spectrum.
As we can see in the table, all four methods, when suitably tuned, agree on the persistence of the process up to fluctuations of $\sim 0.05$ in the Hurst exponent values.
Such values are also similar to the coefficient $H=0.73$ reported in Ref. \cite{hu2009multifractal}, where an adaptive detrending is performed on the time series relative to monthly sunspot.
The Hurst exponent retrieved with DESN, with an ensemble of $k=10$ ESNs, is closer to the ground truth with respect to the other methods, while the outcome obtained with $k=30$ is slightly higher. By assuming that the true value lies in-between the general consensus, this may suggest that a suitable dimension of the ESN ensemble has to be chosen in order to obtain best performance, even if the observed variability is in general fairly low. Regarding the MFW, we observe that DESN is not in agreement with the other detrending methods and, to a lower extent, also on the asymmetry $\Theta$.
In fact, even if all methods agree on the right-sided multifractal nature of the series, both DESN configurations denote a lower degree of multifractality and lower asymmetry.
However, it is worth noting that the MFW value estimated by DESN is much closer to the values reported in \cite{drozdz2015detecting}, while the degree of asymmetry is still different. It is also worth pointing out that the authors in \cite{drozdz2015detecting} did not perform any detrending in their work. This was possible thanks to the fact that the underlying trend is very slow and a number of sufficient data points can be analyzed by considering scales lower than half of the dominating periodicity.
In Fig.~\ref{fig:trendSun}, we show the trends identified using the different approaches herein taken into account. As it is possible to observe, the trend calculated by DESN correctly recognizes the characteristic 11-year cycle of the sunspot time series. 
In Fig.~\ref{fig:scalingSun}, we show the results of the scaling of the fluctuation function obtained by using the two configurations for $k$ of DESN.
The general agreement of the values estimated by DESN with other methods offers a sound justification for the quality and reliability of the proposed detrending method.

\bgroup
\def\arraystretch{1.5} 
\setlength\tabcolsep{1em} 
\begin{table}[!ht]\small
\centering
\caption{Hurst exponent, MFW, and asymmetry ($\Theta$) of the detrended sunspot time series, estimated using different detrending methods.}
\vspace{0.3cm}
\begin{tabular}{llll}
\hline
\textbf{Method} & \textbf{Hurst} & \textbf{MFW} & $\mathbf{\Theta}$ \\
\hline
DESN ($k = 10$) & $ 0.729 \pm 0.0003 $ & $ 0.456 \pm 0.0560 $ & $-0.408 \pm 0.0536 $ \\
DESN ($k = 30$) & $ 0.808 \pm 0.0002 $ & $ 0.641 \pm 0.0614 $ &$-0.542 \pm 0.0412 $ \\
FDFA & $0.688$ & $1.205$ & $-0.556$ \\
SM & $ 0.680$ & $ 1.118$ & $-0.726$\\
EMD & $ 0.731 $ & $ 1.686$ & $-0.786$ \\
\hline
\end{tabular}
\label{tab:resultsSunspot}
\end{table}
\egroup

\clearpage
\section{Conclusions}
\label{sec:conclusions}

In this paper, we have explored the possibility of identifying and removing trends in a given time series by means of echo state networks, a particular type of recurrent neural network.
The proposed method, called DESN, allows to filter out trends with minimal assumptions and without performing a windowed fitting as proposed in other detrending approaches.
This is possible by exploiting the capability of recurrent neural networks to learn and predict complex dynamical processes in order to separate the actual trend from its stochastic fluctuations.
Our main assumption consists in considering the noise and trends components as processes with very different degrees of predictability. We exploited such an assumption as a separating criterion.
Notably, we have used an ensemble of echo state networks as a filter, operating with a standard configuration and trained using linear regression for the readout layer.
Many other approaches exist both for designing the reservoir and for training the readout \cite{dutoit2009pruning,scardapane2014effective}, which could be evaluated in future works depending on the specific problem at hand.

As a first benchmark, we have analyzed the performance of DESN and other detrending techniques taken from the literature on several synthetic time series generated using different types of trends and noise processes.
The quality of the detrending has been evaluated by comparing the properties of the estimated noise with respect to the known ground truths. The evaluations of the Hurst exponents and the properties of the multifractal spectra on the detrended series have been performed with the multifractal detrended fluctuation analysis procedure, a consolidated method in the field of fractal analysis of time series.
In most cases, the resulting fractal coefficients computed by DESN procedure agreed with the expected values and the noise self-similarity properties were preserved by the detrending operation.
On the other hand, in several occasions other detrending methods were not able to perform a correct detrending, which resulted in an incorrect scaling of the fluctuation function.
In general, DESN and a detrending method based on Fourier analysis have shown to be the most reliable methods in terms of detrending accuracy on the considered synthetic time series. 

As a second test, we have analyzed the well-known sunspot time series, which is a multifractal time series that has been taken into account in several related works \cite{drozdz2015detecting,hu2009multifractal}.
Our experimental results suggest that the multifractal properties retrieved by using DESN were both qualitatively and quantitatively compatible with those suggested in other works taken from the literature.
This further strengthens the validity of the proposed data-driven detrending method based on echo state networks.

\bibliographystyle{abbrvnat}
\bibliography{Bibliography}

\end{document}